\documentclass[twocolumn,showpacs,preprintnumbers,amsmath,amssymb]{revtex4}

\usepackage{graphicx}
\usepackage{dcolumn}
\usepackage{bm}

\begin{document}


\title{Wigner-Yanase skew information as tests
for quantum entanglement}
\author{Zeqian Chen}
\email{zqchen@wipm.ac.cn}
\affiliation{%
Wuhan Institute of Physics and Mathematics, Chinese Academy of
Sciences, P.O.Box 71010, 30 West District, Xiao-Hong-Shan,
Wuhan 430071, China}%

\date{\today}

\begin{abstract}
A Bell-type inequality is proposed in terms of Wigner-Yanase skew
information, which is quadratic and involves only one local spin
observable at each site. This inequality presents a hierarchic
classification of all states of multipartite quantum systems from
separable to fully entangled states, which is more powerful than
the one presented by quadratic Bell inequalities from
two-entangled to fully entangled states. In particular, it is
proved that the inequality provides an exact test to distinguish
entangled from nonentangled pure states of two qubits. Our
inequality sheds considerable light on relationships between
quantum entanglement and information theory.
\end{abstract}

\maketitle

As is well known, entangled states have become a key concept in
quantum mechanics nowadays. On the other hand, from a practical
point of view entangled states have found numerous applications in
quantum information [1]. A natural question is then how to detect
entangled states. The first efficient tool used to detect
entangled states is the Bell inequality [2], which was originally
designed to rule out various kinds of local hidden variable
theories. Precisely, the Bell inequality indicates that certain
statistical correlations predicted by quantum mechanics for
measurements on two-qubit ensembles cannot be understood within a
realistic picture based on the local realism of Einstein,
Podolsky, and Rosen [3]. However, Gisin's theorem [4] asserts that
all entangled two-qubit pure states violate the
Clauser-Horne-Shimony-Holt (CHSH) inequality [5] for some choice
of spin observables. There are various Bell-type inequalities used
to detect entangled states [6]. In particular, quadratic Bell-type
inequalities were derived by Uffink [7] as tests for multipartite
entanglement and used to classify all states of $n$ qubits into
$n-1$ entanglement classes from two-entangled to $n$-entangled
(fully entangled) states [8]. Although there are some other
measures of entangled states [1], the problem of how to classify
and quantify entanglement in general is still far from being
completely understood today.

In this article, a Bell-type inequality is proposed in terms of
Wigner-Yanase skew information, which is shown to be useful for
detecting entangled states. In the study of measurement theory
[9], Wigner and Yanase [10] introduced the following quantity from
an information-theoretic point of view\begin{equation}I(\rho, A) =
- \frac{1}{2}tr \left [ \rho^{1/2}, A \right
]^2,\end{equation}where $\rho$ is a state and $A$ an observable of
a quantum system. This quantity $I(\rho, A)$ is called
Wigner-Yanase skew information; it is the amount of information on
the values of observables not commuting with (being skew to) $A.$
They proved that this quantity satisfies the requirements that an
expression for the information content should satisfy, among which
two basic ones are the following.

(a) When two different ensembles are united, the skew information
decreases. Phrased mathematically, $I(\rho, A)$ is a convex
function in $\rho$---if $\rho_1$ and $\rho_2$ are both density
operators, then\begin{equation}I(\alpha \rho_1 + \beta \rho_2, A)
\leq \alpha I(\rho_1, A) + \beta I(\rho_2, A),
\end{equation}where $\alpha + \beta =1$ for $\alpha, \beta \geq
0.$

(b) The skew information content of the union of two independent
systems is the sum of the skew information content of the
individuals; namely, let $\rho_1$ and $\rho_2$ be two states of
the first and second systems, respectively, and $A_1$ and $A_2$
the corresponding local observables; then\begin{equation}I( \rho_1
\otimes \rho_2, A_1 \otimes 1 + 1 \otimes A_2) = I(\rho_1, A_1) +
I(\rho_2, A_2).\end{equation}The equality (3) can be extended to
several independent systems.

There are some earlier works concerning the Wigner-Yanase skew
information. These constitute the celebrated Wigner-Araki-Yanase
theorem, which puts a limitation on the measurement of observables
in the presence of a conserved quantity [11], and was further
investigated by many authors [12]. Recently, Luo and Zhang [13]
reveal that the Wigner-Yanase skew information can be regarded as
the quantum analog of the Fisher information in the theory of
statistical estimation, and used to establish some results
concerning the characterizations of uncertainty relations and
measures of entangled states. Here, we present a Bell-type
inequality of Wigner-Yanase skew information for separable states
of $n$ particles.

First of all, note that the Wigner-Yanase skew information can be
rewritten as\begin{equation}I(\rho, A) = tr \rho A^2 - tr
\rho^{1/2} A \rho^{1/2} A.\end{equation}In particular, if $\rho =
|\psi \rangle \langle \psi |$ is a pure state,
then\begin{equation}I(| \psi \rangle, A) = \langle \psi |A^2| \psi
\rangle - \langle \psi |A| \psi \rangle^2.\end{equation}From
Eqs.(2) and (5) one concludes that\begin{equation}I(\rho, A) \leq
1 \end{equation}for each spin observable $A$ ($A^2 = 1$) in all
states $\rho.$

Now, consider a system of $n$ particles and assume that a local
spin observable is measured on each particle, and hence there
correspond $n$ local spin observables $A_1,\ldots, A_n.$ Since
separable states are written as a convex sum over the states
$\rho_1 \otimes\cdots \otimes \rho_n,$ and by Eqs.(3) and (6), one
has$$I(\rho_1 \otimes\cdots \otimes \rho_n, A_1 + \cdots + A_n) =
\sum^n_{j=1} I(\rho_j, A_j) \leq n,$$it is concluded from Eq.(2)
that\begin{equation}I(\rho, A_1 + \cdots + A_n) \leq n
\end{equation}for all separable states $\rho$ of the $n$ particles.
(We write $A_1,$ etc., as shorthand for $A_1\otimes 1 \otimes
\cdots \otimes 1.$) The inequality (7) is a Bell-type inequality
in terms of Wigner-Yanase skew information, which can be violated
by entangled states. We will show that this inequality not only
distinguishes, for systems of $n$ particles, between fully
entangled states and states in which at most $n-1$ particles are
entangled, similar to quadratic Bell inequalities [7], but also
presents a $n$-hierarchic classification of all states of
$n$-partite quantum systems from one-entangled (separable) to
$n$-entangled (fully entangled) states, which is more powerful
than the one presented by quadratic Bell inequalities giving $n-1$
entanglement classes from two-entangled to fully entangled states
[8]. Roughly speaking, the more entangled the larger the violation
value of Eq.(7), which means it is easier to measure an observable
not commuting with local spin observables on entangled states,
thus corresponding to larger skew information. This directly
uncovers the information-theoretic meaning of entangled states.

We would like to mention that, contrary to the usual Bell-type
inequalities possessing two local spin observables or more at each
site [14], Eq.(7) involves only one local spin observable at each
site. On the other hand, our inequality, similarly to quadratic
Bell-type inequalities [7], is not linear but quadratic. However,
we will show that Gisin's theorem holds true for Eq.(7), that is,
it provides an exact test to distinguish entangled from
nonentangled pure states of two qubits. This shows that our
inequality is stronger than quadratic Bell-type inequalities,
which provide tests for entanglement for systems of only three
particles or more, and not for $n=2.$

Since $I(\rho, A)$ is a convex function in $\rho,$ the maximal
violation of Eq.(7) for entangled states will occur in pure
states. Note that for local spin observables $A_1,\ldots,
A_n,$$$\left \langle \psi \left |(A_1 + \cdots + A_n )^2 \right |
\psi \right \rangle = \sum_{j,k} \left \langle \psi \left | A_jA_k
\right | \psi \right \rangle \leq n^2,$$($A_jA_k = 1\otimes\cdots
\otimes A_j \otimes \cdots \otimes A_k \otimes \cdots \otimes 1$
in short) we conclude from Eq.(5) that\begin{equation}I(\rho, A_1
+ \cdots + A_n) \leq n^2\end{equation}for all (entangled) states
$\rho$ of the $n$ particles. Equality in Eq.(8) can be attained
even for systems of $n$ spin-$\frac{1}{2}$ particles by the
Greenberger-Horne-Zeilinger (GHZ) state [15]$$|GHZ \rangle =
\frac{1}{\sqrt{2}} \left ( |0\cdots 0 \rangle + |1 \cdots 1
\rangle \right ).$$Indeed, choosing $A_j = \sigma^j_z$
($j=1,\ldots, n$) one has that$$\sigma^j_z \sigma^k_z | GHZ
\rangle = | GHZ \rangle, j,k=1,\ldots, n,$$$$\langle GHZ
|\sigma^1_z + \cdots + \sigma^n_z | GHZ \rangle =0.$$Hence, by
Eq.(5) one has\begin{equation}I(| GHZ \rangle, \sigma^1_z + \cdots
+ \sigma^n_z) = n^2.\end{equation}

In particular, for the familiar case of two spin-$\frac{1}{2}$
particles, the maximal violation of Eq.(7) occurs for Bell states
with the violation value $4,$ which is larger than the maximal
violation value $2\sqrt{2}$ of the CHSH inequality [16]. Moreover,
for a pure state $|\psi \rangle$ of two qubits we can write by the
Schmidt decomposition theorem\begin{equation}|\psi \rangle = p |
\phi_1\rangle|\chi_1\rangle + q
|\phi_2\rangle|\chi_2\rangle,\end{equation}where $p$ and $q$ are
two nonnegative numbers satisfying $p^2 + q^2 =1$ and
$\{|\phi_1\rangle,|\phi_2\rangle\}$ and $\{|\chi_1\rangle,
|\chi_2\rangle \}$ are orthonormal bases in the two-dimensional
Hilbert spaces ${\cal H}_1$ and ${\cal H}_2$ of the two particles,
respectively. Now, choose a representation for the first particle
such that $|\phi_1\rangle = |0 \rangle_1,|\phi_2\rangle = |1
\rangle_1,$ in the $z_1$ direction, and similarly, $|\chi_1\rangle
= |0 \rangle_2,|\chi_2\rangle = |1 \rangle_2,$ in the $z_2$
direction. Then, a simple computation yields
that\begin{equation}I(|\psi \rangle, \sigma^1_x + \sigma^2_x) = 2
+ 4 pq.\end{equation}This shows that $|\psi \rangle$ is entangled
($pq > 0$) if and only if it violates Eq.(7), that is, Gisin's
theorem holds true for the inequality (7) of two qubits.
Therefore, Wigner-Yanase skew information provides an exact test
to distinguish entangled from nonentangled pure states of two
qubits.

In the following, we show how to detect multipartite entanglement
using Wigner-Yanase skew information . Consider a system of $n$
particles. Recall that a state of $n$ particles is fully entangled
[17], if it cannot be written as $\rho_{(k)} \otimes \rho_{(n-k)}$
or mixtures of these states, where $\rho_{(k)}$ ($1 \leq k \leq
n-1$) is a (entangled or not) state of $k$ particles among the $n$
particles. For example, states of three particles that are not of
the form $\rho_1 \otimes \rho_{23},$ $\rho_2 \otimes \rho_{13},$
$\rho_{12} \otimes \rho_3,$ or mixtures of these states are fully
entangled states. In particular, $|GHZ \rangle$ are fully
entangled states of multiqubit systems. Generally, for $1 \leq k
\leq n,$ a state of the $n$ particles is at most $k$-entangled, if
it is of the form $\rho_{(k_1)} \otimes \cdots \otimes
\rho_{(k_m)}$ or mixtures of these states, where $1 \leq k_j \leq
k,$ $\sum_j k_j = n,$ and $\rho_{(k_j)}$ is a (entangled or not)
state of $k_j$ particles among the $n$ particles. We denote by
$ES_k$ all states that are at most $k$-entangled.
Then,\begin{equation}ES_1 \subset ES_2 \subset \cdots \subset
ES_n.\end{equation}Clearly, $ES_1$ is the set of all separable
states, $ES_n$ are all states of the $n$ particles, and $ES_n -
ES_{n-1}$ are all fully entangled states. Thus, for $2 \leq k \leq
n,$ $k$-entangled states are all states in $ES_k - ES_{k-1},$
$1$-entangled states are separable states, and $n$-entangled
states are just fully entangled states.

Given a state of the form $\rho_{(k_1)} \otimes \cdots \otimes
\rho_{(k_m)}.$ For local spin observables $A_1,\ldots, A_n,$ by
Eqs.(3) and (8) we have$$\begin{array}{l}I(\rho_{(k_1)} \otimes
\cdots \otimes \rho_{(k_m)}, A_1 + \cdots + A_n )\\ =
I(\rho_{(k_1)}, A_{(k_1)}) + \cdots + I(\rho_{(k_m)}, A_{(k_m)})\\
\leq k^2_1 +\cdots + k^2_m,\end{array}$$where $A_{(k_j)}$ is the
sum of the corresponding local spin observables among $A_1,\ldots,
A_n$ associated with $\rho_{(k_j)}.$ Then, by the convexity (a),
we obtain\begin{equation}I(\rho, A_1 + \cdots + A_n )\leq \left [
\frac{n}{k} \right ] k^2 + \left ( n - \left [ \frac{n}{k} \right
] k \right )^2\end{equation}for all states $\rho \in ES_k,$ where
$[x]$ denotes the largest integer less than or equal to $x.$ Here,
the right-hand side of Eq.(13)$$E_k = \left [ \frac{n}{k} \right ]
k^2 + \left ( n - \left [ \frac{n}{k} \right ] k \right )^2$$can
be obtained by induction [18]. Clearly,\begin{equation}E_1 = n <
E_2 <\cdots < E_n = n^2.\end{equation}Also, $E_2 = 2n$ for $n$
even or $2n-1$ for $n$ odd and, $E_{n-1} = (n-1)^2 + 1.$ For
example, in the cases of $n=2,3, 4,$ and $5,$ we have
\begin{description}
\item{(1)} $n=2:$ $E_1 =2$ and $E_2 = 4;$ \item{(2)} $n=3:$ $E_1 =
3, E_2 = 5,$ and $E_3 = 9;$ \item{(3)} $n=4:$ $E_1 =4, E_2 = 8,
E_3 = 10, E_4 = 16;$ and\item{(4)} $n=5:$ $E_1 =5, E_2 = 9, E_3 =
13, E_4 = 17, E_5 =25.$\end{description}Moreover, $E_k$ can be
attained in Eq.(13) by $k$-entangled states. Indeed, choosing
$\rho = \rho_{(k_1)} \otimes \cdots \otimes \rho_{(k_m)}$ such
that $m= \left [ \frac{n}{k} \right ] + 1, k_j = k$ for
$j=1,\ldots, \left [ \frac{n}{k} \right ],$ $\rho_{(k_j)}$ is the
GHZ state of the corresponding $k$ particles, and $\rho_{(k_m)}$
is the GHZ state of the remaining particles, by the additivity (b)
and Eq.(9) we have$$I(\rho, \sigma^1_z + \cdots + \sigma^n_z) =
E_k.$$Note that the inequality (13) is only a necessary condition
that states are at most $k$-entangled. Later, we will present a
concrete example showing that there are fully entangled states
satisfying Eq.(7).

Now, we turn to the problem of how to classify states of $n$
particles into $n$ entanglement classes by using Wigner-Yanase
skew information. To this end, we define\begin{equation}I(\rho) =
\sup I (\rho, A_1 + \cdots + A_n),\end{equation}where the supremum
is taken over all local spin observables $A_1,  \cdots,  A_n.$
According to the physical meaning of Wigner-Yanase skew
information, the quantity $I(\rho)$ is the most amount of
information on the values of observables not commuting with (being
skew to) local spin observables we can obtain on the state $\rho.$
Hence, $I(\rho)$ can be regarded as the nonlocal skew information
of $\rho.$ By Eq.(13) we find that the larger the nonlocal skew
information, the more entangled or the less separable is the
state. This means that we can obtain more information of some
nonlocal observables on entangled states than on separable ones.
In this sense, we may regard entangled states as an
information-theoretic concept. In particular, by using the
nonlocal skew information of states of multipartite systems we
obtain a classification of them as follows.

{\it Classification theorem}. The states of $n$ particles can be
classified into $n$ entanglement classes. For $k=1,\ldots, n,$ the
states in the $k$th entanglement class
satisfy\begin{equation}I(\rho) \leq E_k.\end{equation}Therefore,
the larger the nonlocal skew information, the more entangled or
the less separable is the state in the sense that a larger maximal
violation of the inequality $(13)$ is attainable for this class of
states.

As shown in Eq.(13), $k$-entangled states are all in the $k$th
entanglement class. In particular, if$$I(\rho)> (n-1)^2 + 1,$$then
$\rho$ is fully entangled. For example, Eq.(9) shows that $|GHZ
\rangle$ is a fully entangled state. Hence, by using Wigner-Yanase
skew information, we give a hierarchic classification of
multipartite entangled states.

For each separable pure state
$|\psi_1\rangle\cdots|\psi_n\rangle,$ it is easy to check that
$I(|\psi_1\rangle\cdots|\psi_n\rangle)= n.$ For mixed states, we
illustrate the following states of $n$ qubits
[17]\begin{equation}\rho_{\lambda} = \lambda |GHZ \rangle\langle
GHZ | + \frac{1- \lambda}{2^n} I,\end{equation}where $0 \leq
\lambda \leq 1$ and $I$ is the identity on the $n$ qubits.
Since$$\rho^{1/2}_{\lambda} = f(\lambda ) |GHZ \rangle\langle GHZ|
+ \sqrt{\frac{1-\lambda}{2^n}}I,$$where $f(\lambda ) =
\sqrt{\lambda + \frac{1-\lambda}{2^n}} -
\sqrt{\frac{1-\lambda}{2^n}},$ we
have$$\begin{array}{lcl}I(\rho_{\lambda}, A) &=& \left ( \lambda -
2 f(\lambda )\sqrt{\frac{1-\lambda}{2^n}} \right ) \langle GHZ|
A^2
|GHZ \rangle\\&~& - f(\lambda )^2 \langle GHZ |A| GHZ\rangle^2\\
&\leq& \left ( \lambda - 2 f(\lambda )\sqrt{\frac{1-\lambda}{2^n}}
\right ) n^2,\end{array}$$for all $A = A_1 + \cdots + A_n$ of
local spin observables. Taking $A_j = \sigma^j_z$ for $j=1,\ldots,
n,$ we have\begin{equation}I(\rho_{\lambda}) = \left [ \lambda -
2\sqrt{\frac{1-\lambda}{2^n}}\left ( \sqrt{\lambda +
\frac{1-\lambda}{2^n}} - \sqrt{\frac{1-\lambda}{2^n}} \right )
\right ]n^2.\end{equation} It is shown in [17] that
$\rho_{\lambda}$ is separable if and only if $0 \leq \lambda \leq
1/(1+2^{n-1} ),$ and fully entangled otherwise. The lower bound of
values $\lambda$ for which our criterion can be used to detect the
(full) entanglement of $\rho_{\lambda}$
is\begin{equation}\lambda_n = \frac{1}{n}\left ( 1-
\frac{1}{2^{n-1}} \right ) + \sqrt{\frac{1}{2^{n-2}n} \left ( 1 -
\frac{1}{n} + \frac{1}{2^nn} \right )},\end{equation}that is,
$I(\rho_{\lambda}) > n$ provided $\lambda > \lambda_n.$ We
have\begin{description} \item{(i)} $\lambda_2 =
\frac{1+\sqrt{5}}{4}$ and $\lambda_3 = \frac{3+\sqrt{17}}{12};$
\item{(ii)} $\lambda_4 = \frac{7}{16} < \frac{1}{2}$ and
$\lambda_5 = \frac{15+ \sqrt{129}}{80} < \frac{1}{3};$
\item{(iii)} $\lambda_6 = \frac{31+ \sqrt{321}}{192} <
\frac{1}{3}$ and $\lambda_7 = \frac{63+ \sqrt{769}}{448} <
\frac{1}{4} .$\end{description}Moreover, we have the following
estimates\begin{equation}\frac{1}{n-1} < \lambda_n < \frac{1}{n-2}
,~~8 \leq n \leq 12;\end{equation}
\begin{equation}\frac{1}{n} < \lambda_n <
\frac{1}{n-1},~~n> 12.\end{equation}On the other hand, it is easy
to check that the lower bound of values $\lambda$ is $1/2^{n/2-1}$
for the (full) entanglement of $\rho_{\lambda}$ obtained by the
quadratic Bell inequality [7], as well as $1/2^{(n-1)/2}$ by the
standard Mermin-Klyshko inequality [6,14]. Although the lower
bounds of values derived by the quadratic Bell and Mermin-Klyshko
inequalities, respectively, are both better than the one of our
inequality, our criterion, which requires only one measurement
setting per site, is more appealing from an experimental point of
view.

Setting $\lambda_0 = 1/(1+2^{n-1} ),$ one has$$I(\rho_{\lambda_0})
= \frac{2-\sqrt{3}}{1+ 2^{n-1}}n^2.$$Since
$I(\rho_{\lambda_0})\longrightarrow 0$ as $n\longrightarrow
\infty$ and by Eq.(18), $I(\rho_{\lambda})$ is continuous in
$\lambda,$ we find that there are some fully entangled states
$\rho_{\lambda}$ with $\lambda > 1/( 1+2^{n-1} )$ such that
$I(\rho_{\lambda})\longrightarrow 0$ as $n\longrightarrow \infty.$
This means that the nonlocal skew information of some fully
entangled states can be arbitrarily small and hence smaller than
that of separable states. Since $I(\rho)$ can be regarded as a
measure of nonlocality of quantum states, therefore, from the
information-theoretic viewpoint there is a gap between nonlocality
and quantum entanglement.

As follows, we apply our theorem to the classification of
three-qubit states. As stated before, there are three types of
three-qubit states: one-entangled (totally separable) states
denoted by $ES_1,$ two-entangled states $ES_2,$ and
three-entangled (fully entangled) states $ES_3.$ The results for
the Mermin-Klyshko (MK) inequalities [6], the quadratic Bell
inequality (BI$_2$) [7], and our inequality of Wigner-Yanase (WY)
information are summarized in Table I.\begin{table}
\caption{\label{tab:table1}Each row indicates the maximum values
under three different classes possibly attained by MK, BI$_2,$ and
WY, respectively. Note that the difference values of WY
two-entangled from separable and fully entangled from
two-entangled states are 2 and 4 respectively, which are both
larger than the associated ones $\sqrt{2}-1$ and $2-\sqrt{2}$ of
MK.}
\begin{ruledtabular}
\begin{tabular}{cccc}
 & $ES_1$ & $ES_2$ & $ES_3$\\
\hline MK & 1 & $\sqrt{2}$ & 2\\
BI$_2$& 8 & 8 & 16\\
WY& 3 & 5 & 9\\
\end{tabular}
\end{ruledtabular}
\end{table}

We now turn to our entanglement criterion to see how good it is to
test the entanglement of generalized GHZ states of three
qubits$$|\psi \rangle = \alpha |000 \rangle + \beta |111
\rangle,$$where $\alpha , \beta > 0,$ and $\alpha^2 + \beta^2 =1.$
Write $A_j = a_{j1} \sigma_x + a_{j2} \sigma_y + a_{j3} \sigma_z,$
where $\vec{a}_j = (a_{j1}, a_{j2}, a_{j3} ) \in \mathbb{R}^3$ are
all unit vectors. Note that$$\begin{array}{l}I(|\psi \rangle,
A_1+A_2+A_3)\\= 3+ 2( a_{13} a_{23} + a_{13} a_{33} + a_{23}
a_{33})\\~~~~- (\alpha^2 - \beta^2)^2 (a_{13} + a_{23} +
a_{33})^2.\end{array}$$Hence, we have\begin{equation}I(|\psi
\rangle) = 3 + 3[2-3(\alpha^2 - \beta^2)^2].\end{equation}It is
concluded that whenever
\begin{equation}\alpha, \beta
> \sqrt{\frac{1}{2}\left ( 1- \sqrt{\frac{2}{3}} \right )},\end{equation}
our criterion can detect the entanglement of generalized GHZ
states, which is more efficient than the MK inequality with the
critical value $\sqrt{( 1- 1/ \sqrt{2} )/2}.$

In summary, a Bell-type inequality is proposed in terms of
Wigner-Yanase skew information, which presents an $n$-hierarchic
classification of all states of $n$-partite quantum systems from
separable to fully entangled states. The inequality is not linear
but quadratic; however, contrary to quadratic Bell inequalities
[7], it involves only one local spin observable at each site and
provides an exact test to distinguish entangled from nonentangled
pure states of two qubits. Our $n$-hierarchic classification for
$n$-partite entangled states by using Wigner-Yanase skew
information is more powerful than the classification in terms of
quadratic Bell inequalities [8], which only classifies all states
of $n$ particles into $n-1$ entanglement classes from
two-entangled to $n$-entangled states and cannot be used to
distinguish two-entangled from separable states. We have defined
the quantity $I(\rho )$ of multipartite states, which is the most
amount of information on the values of observables not commuting
with (being skew to) local spin observables we can obtain on the
state $\rho.$ $I(\rho)$ is the nonlocal skew information of
entangled states, as we can obtain more information of some
nonlocal observables on entangled states than on separable ones.

Nowadays, it has been recognized that most physical processes in
nature can be formulated in terms of processing of information,
and information may be central to understanding quantum theory
[19]. The notion of the Wigner-Yanase skew information quantifies
the amount of information on the values of observables being skew
to other ones [11,12], and has been used to establish some results
concerning the characterizations of uncertainty relations and
measures of entangled states [13]. Our results furthermore shed
considerable light on relationships between quantum entanglement
and information theory in terms of Bell-type inequalities of the
Wigner-Yanase skew information. We expect that, similarly to the
usual Bell inequalities [6], the Bell-type inequality of the
Wigner-Yanase skew information and its various ramifications will
play an important role in quantum information [1].

This work was partially supported by the 973 Project of China
(Grant No. 2001CB3093).

\newpage 


\begin{thebibliography}{**}
\bibitem{1}N.Gisin, G.Ribordy, W.Tittle, and H.Zbinden,
Rev. Mod. Phys. {\bf 74}, 145(2002); M.Nielsen and I.Chuang, {\it
Quantum Computation and Quantum Information} (Cambridge University
Press, Cambridge, England, 2000).
\bibitem{2}J.S.Bell, Physics (Long Island City, N.Y.){\bf 1},
195(1964).
\bibitem{3}A.Einstein, B.Podolsky, and N.Rosen, Phys. Rev. {\bf
47}, 777(1935).
\bibitem{4}N.Gisin, Phys.Lett. A {\bf 154}, 201(1991).
\bibitem{5}J.F.Clauser, M.A.Horne, A.Shimony, R.A.Holt,
Phys.Rev.Lett. {\bf 23}, 880(1969).
\bibitem{6}D.Collins, N.Gisin, S.Popescu, D.Roberts, and
V.Scarani, Phys.Rev.Lett.{\bf 88}, 170405(2002); M.Seevinck and
G.Svetlichny, Phys.Rev.Lett.{\bf 89}, 060401(2002). More
references can be found in e-print quant-ph.
\bibitem{7}J.Uffink, Phys.Rev.Lett.{\bf 88}, 230406(2002);
K.Nagata, M.Koashi, and N.Imoto, Phys.Rev.Lett.{\bf 89},
260401(2002).
\bibitem{8}S.-X.Yu, Z.-B.Chen, J.-W.Pan, and Y.-D.Zhang,
Phys.Rev.Lett.{\bf 90}, 080401(2003).
\bibitem{9}E.P.Wigner, Z.Phys.{\bf 133}, 101(1952);
E.P.Wigner, {\it Physikertagung Wien} (Physik Verlag,
Mosbach/Baden, 1952), p.1.
\bibitem{10}E.P.Wigner and M.M.Yanase, Proc.Natl.Acad.Sci.U.S.A.
{\bf 49}, 910(1963).
\bibitem{11}H.Araki and M.M.Yanase, Phys.Rev.{\bf 120}, 622(1960);
M.M.Yanase, Phys.Rev.{\bf 123}, 666(1961).
\bibitem{12}M.Ozawa, Phys.Rev.Lett. {\bf 67}, 1956(1991); Phys.Rev.Lett. {\bf 88},
050402(2002); Phys.Rev.Lett. {\bf 89}, 057902(2002); S.Matsumoto,
Prog.Theor.Phys.{\bf 90}, 35(1993); K.Kakazu and S.Pascazio,
Phys.Rev. A {\bf 51}, 3469(1995).
\bibitem{13}S.-L.Luo, Phys.Rev.Lett.{\bf 91}, 180403(2003);
S.-L.Luo and Q.Zhang, Phys.Rev. A {\bf 69}, 032106(2004).
\bibitem{14}R.F.Werner and M.M.Wolf, Phys.Rev. A {\bf 64}, 032112(2001);
M.\.{Z}ukowski and \v{C}.Brukner, Phys.Rev.Lett.{\bf 88},
210401(2002).
\bibitem{15}D.M.Greenberger, M.A.Horne, and A.Zeilinger, in {\it
Bell's Theorem, Quantum Theory, and Conceptions of the Universe,}
edited by M.Kafatos (Kluwer, Dordrecht, 1989), p.69; N.D.Mermin,
Phys. Today {\bf 43}, No. 6, 9(1990); N.D.Mermin, Am. J. Phys.
{\bf 58}, 731(1990).
\bibitem{16}B.S.Cirel'son, Lett.Math.Phys. {\bf 4}, 93(1980).
\bibitem{17}W.D\"{u}r and J.I.Cirac, Phys.Rev. A {\bf 61},
042314(2000).
\bibitem{18}Consider the function $f = x^2_1 + \cdots + x^2_n$ on
the convex set $C_k = \{ (x_1,\ldots, x_n): 0 \leq x_j \leq k,
\sum_j x_j = n \}.$ Since $f$ is strictly convex, it attains the
maximum at an extreme point of the convex set $C_k.$ Then, without
loss of generality, assuming $x_1 \geq x_2 \geq \cdots \geq x_n,$
we have that $x_1 = k.$ Inductively, we obtain Eq.(13).
\bibitem{19}J.Wheeler, in {\it Complexity, Entropy, and Physics of
Information,} edited by Z.H.Zurek (Addison-Wesley, Reading, MA,
1990), pp.3-28; J.Summhammer, Int.J.Theor.Phys. {\bf 33},
171(1994); B.R.Frieden, {\it Science from Fisher Information: A
Unification}(Cambridge University Press, Cambridge, England,
2004).
\end{thebibliography}
\end{document}